\documentclass[%
 reprint,
 amsmath,amssymb,
 aps,
]{revtex4-2}

\usepackage{graphicx}% Include figure files
\usepackage{dcolumn}% Align table columns on decimal point
\usepackage{bm}% bold math
\usepackage{tikz}
\usetikzlibrary{patterns}
\usetikzlibrary{shapes.geometric}
\usepackage[version=4]{mhchem}
\begin{document}

\preprint{APS/123-QED}

\title{Detection and Ranging Beyond the Canonical Resolution Limit}

\author{Nathaniel J. Fuller}
 \email{nathaniel.j.fuller3.civ@us.navy.mil}
\author{Nicholas Palermo}
 \email{nicholas.a.palermo2.civ@us.navy.mil}
\affiliation{Naval Surface Warfare Ctr. - Panama City Div.}

\date{\today}

\begin{abstract}
The canonical range resolution limit in radar, sonar, and lidar systems is found to be a special case of a more general resolution limit. The general limit indicates that it is possible to surpass the canonical limit in moderate (of order unity) signal-to-noise ratio (SNR) environments by using the signal amplitude and phase information. The canonical limit only considers the bandwidth of the received signal without considering how SNR affects the range resolution. Details present in the signal amplitude, such as attenuation and geometric spreading, can act as additional sources of range information. Previous studies have taken advantage of the relationship between target distance and signal amplitude or phase to achieve higher resolution ranging, and often employ unusual transmit waveforms for this purpose. These methods each provide distinct bounds on range resolution, rather than a unified bound applicable across different systems and applications. We apply ideas from information theory to determine a general lower bound to the smallest resolvable range bin size and corresponding target strength measurements.
\end{abstract}

%\keywords{Suggested keywords}%Use showkeys class option if keyword
                              %display desired

\maketitle

\section{Introduction}
Typical radar, sonar, and lidar systems share limitations in their range resolution that are controlled by the bandwidth of the transmit pulse. This \textit{canonical} resolution limit appears to be a special case of a more general limit that also depends on the signal to noise ratio (SNR) of the reflected pulse. The general limit indicates that it is possible to surpass the canonical limit in a moderate SNR environment by using signal processing techniques which do not rely solely on the output of a matched filter for ranging. Details present in the signal amplitude, such as attenuation and geometric spreading, and in the signal phase, can act as additional sources of range information. Previous studies \cite{Wong2012, ko2008range, Komissarov2019, Howell2023, borison1992super, bajwa2011identification, suwa2002bandwidth} have taken advantage of these information sources to achieve higher resolution ranging. These methods often use exotic waveforms and seem to have different resolution bounds depending on the approach used. We apply ideas from information theory to determine a general lower bound to the distance between scatterers that can be reliably resolved in range.

\section{Background}
Remote range sensing is accomplished by periodically transmitting a known signal and measuring the return time or return time-dependent effects \cite{brooker2005understanding} of the scattered signal to determine the range of the scatterer. The return signal is typically passed through a matched filter to reject noise in the received signal \cite{Richards2014}. The matched filter output consists of a set intensity values across discrete time bins. (Note that this discrete view is valid for continuous analog signals as well due to the Nyquist–Shannon sampling theorem.) A high intensity bin normally indicates the presence of a scattering surface at the corresponding range. Closely spaced bins are typically desirable to resolve the individual scattering surfaces of an object and determine its shape.  However, the ability to resolve closely spaced scatterers in this way is limited by the bandwidth of the signal. To see this, first note that the total number of time bins equals the duration of the signal (or the time between pings) $T$ times the sample rate. Assuming Nyquist sampling, the sample rate is equal to the bandwidth $B$. The number of time bins is now $N=TB$. Since the maximum range covered is $R=cT/2$ where $c$ is the wave speed, we now have $N=2BR/c$. The range resolution $\delta$ is the range bin width $R/N$, giving the canonical resolution limit \cite{skolnik1960theoretical},
\begin{equation}
\delta\geq \frac{c}{2B}.
\label{canon_res_limit}
\end{equation}
\par
This limit has led to the development and use of various technologies to increase available bandwidth, including sonar systems operating up into the megahertz range \cite{Groen2006, Park2018, Kim2020}, radar systems approaching 80 gigahertz range \cite{ng2014fully}, and radar systems that synthesize a broadband return from multiple narrow-band coherent signals \cite{xue2024range}. However, this overall approach leads to severe attenuation of the transmitted and reflected pulse since attenuation is an increasing function of frequency \cite{Urick1983} and results in shorter operational ranges. Recently, some researchers have reported achieving range resolution beyond the canonical limit \cite{Wong2012, ko2008range, borison1992super, suwa2002bandwidth}, and in some cases by multiple orders of magnitude \cite{Komissarov2019, Howell2023}. These new approaches take advantage of phase and amplitude information present in the signal to further localize individual scatterers by using specialized transmit waveforms and signal processing techniques.

\par
Using the amplitude information to further localize scatterers is reasonable if the return signal has a high SNR. For example, if we consider a perfect system with no noise and known dissipation factors, it would be possible to determine the range of a single scatterer \textit{exactly} since we would know precisely how the received signal amplitude changes with distance. This is of course complicated by the presence of noise and other unknown parameters in a realistic system. It is not clear how this approach would perform in a less controlled noisy environment. The experiments presented in \cite{Howell2023} used a wire as a wave-guide, which is a fairly clean environment where a high SNR can be readily achieved.
\par
The use of phase information for improved localization was achieved in \cite{Komissarov2019} by showing how bandwidth could be traded for a longer duration signal while still providing the needed ranging information. We are not aware of any methods that use both amplitude and phase information simultaneously for ranging, although we see no reason why such an approach would not be possible. We also note that some machine learning approaches for solving the ranging problem \cite{Li2021, Geiss2020, Donini2022, chang2022rangesrn, Schuessler2023} may use both types of information, but it is hard to tell due to the ``black-box'' nature of these algorithms. In the proceeding sections, we develop a generalization of eq. \ref{canon_res_limit} which does not depend on the form of the transmit signal or the details of how the received signal is processed; giving a resolution limit that applies to all approaches for solving the ranging problem.

\section{Ranging of Scatterers}
Due to transmission loss, the return signal amplitude is a decreasing function with increasing distance and therefore contains useful ranging information. We derive the expression for the transmission loss in the appendix \ref{transmission_loss} since some references such as \cite{Urick1983} appear to give an incorrect expression (the dissipation factor is missing the -1 term). The revised transmission loss is,
\begin{equation}
\text{TL}\left(r\right)=20\log(r)+\alpha\left(r-1\right),
\label{tl}
\end{equation}
where $\alpha$ is the absorption coefficient and controls the exponential decay for both acoustic and electromagnetic waves. The parameter $\alpha$ is also often an increasing function with frequency. This is especially problematic in the sonar domain for high bandwidth systems with high center frequencies where dissipation loss greatly reduces their operational range.
\par
To see how transmission loss is incorporated into the ranging problem, let us first consider the simple case of a small scatterer a distance $r$ from the origin in free space. The scattered wave received by an active sonar system at the origin will be a delayed and attenuated copy of the transmitted signal. The delay $\tau$ is simply the propagation time $\tau=2r/c$. The return signal intensity level $\text{RL}$ in decibels is,
\begin{equation}
\text{RL}\left(r\right)=\text{SL}+\text{TS}-2\text{TL}\left(r\right),
\end{equation}
where $\text{SL}$ is the transmit source level and $\text{TS}$ is the target strength of the scatterer. If a sonar system transmits a known signal $x\left(t\right)$, then the return signal $y\left(t,R\right)$ is,
\begin{equation}
y\left(t,r\right)=10^{(\text{SL}+\text{TS})/20-\text{TL}\left(r\right)/10}e^{i\phi}x\left(t-\tau\right),
\label{atten_eq}
\end{equation}
where $\phi$ represents the phase change from the scatterer which is a function of its material properties.
\par
\begin{figure}
\begin{tikzpicture}
\draw[line width=1,draw=black,fill=blue!20] (1,1) rectangle ++(1.5,0.5);
\draw[line width=1,draw=black,fill=blue!20] (2.5,1) rectangle ++(1.5,0.5);
\draw[dotted,line width=2,draw=gray] (5.5,1.25) -- ++(1.5,0);
\draw[line width=1,draw=black,fill=blue!20] (4,1) rectangle ++(1.5,0.5);
\draw[line width=1,draw=black,fill=blue!20] (7,1) rectangle ++(1.5,0.5);
\draw[] (1,1.75) node[] {$0$};
\draw[] (2.5,1.75) node[] {$\delta_{0}$};
\draw[] (4,1.75) node[] {$2\delta_{0}$};
\draw[] (7,1.75) node[] {$(N-1)\delta_{0}$};
\draw[line width=1,draw=black,fill=blue!20] (7,1) rectangle ++(1.5,0.5);
\node at (1.5, 1.25)[circle,fill,inner sep=1.5pt]{};
\node at (1.9, 1.25)[circle,fill,inner sep=1.5pt]{};
\node at (4.8, 1.25)[circle,fill,inner sep=1.5pt]{};
\node at (8, 1.25)[circle,fill,inner sep=1.5pt]{};
\draw (1,0.7) -- node[below] {$q_{0}$} (1.5, 0.7);
\draw (1,0.6) -- (1, 0.8);
\draw (1.5,0.6) -- (1.5, 0.8);
\draw (2.5,0.7) -- node[below] {$q_{1}$} (1.9, 0.7);
\draw (2.5,0.6) -- (2.5, 0.8);
\draw (1.9,0.6) -- (1.9, 0.8);
\draw (4,0.7) -- node[below] {$q_{2}$} (4.8, 0.7);
\draw (4,0.6) -- (4, 0.8);
\draw (4.8,0.6) -- (4.8, 0.8);
\draw (7,0.7) -- node[below] {$q_{N-1}$} (8, 0.7);
\draw (7,0.6) -- (7, 0.8);
\draw (8,0.6) -- (8, 0.8);
\end{tikzpicture}
\caption{Scatterer locations represented by black dots at locations $q_{n}$ relative to bin positions at $n\delta_{0}$. The offsets $q_{n}$ are depicted without any constraint on their values, allowing scatterers to be placed arbitrarily related to the canonical bin spacing $\delta_{0}$. This can possibly lead to empty bins as shown.}
\label{qbin}
\end{figure}
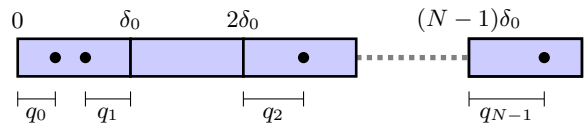

For the more general case of $N$ discrete scatterers at various ranges $r_{n}$, a set of $N$ linearly independent signals must be transmitted to give a sufficient number of degrees of freedom to solve the resulting set of equations for each range. The natural choice is to use the Fourier basis functions $\{\tilde{x}_{m}e^{i\omega_{m}t}\}$ which span the desired bandwidth $B$ given that the downshifted frequency components are $\omega_{m}=2\pi mB/N$. Each basis signal is attenuated and time delayed after returning from each scatterer. The return signal spectral components $\tilde{y}_{m}$ are then,
\begin{equation}
\tilde{y}_{m}=10^{\text{SL}/20}\tilde{x}_{m}\sum_{n=0}^{N-1}a_{n}b_{n}e^{-i\omega_{m}\tau_{n}},
\label{return_spec}
\end{equation}
where $a_{n}=e^{i\phi_{n}}10^{\text{TS}_{n}/20}$ and $b_{n}=10^{\text{-TL}\left(r_{n}\right)/10}$ represent the target strength and transmission loss contributions for the $n$th scatterer. If the scatterer ranges are written as $r_{n}=n\delta_{0}+q_{n}$ and $\delta_{0}=c/2B$, then eq. \ref{return_spec} becomes,
\begin{equation}
H_{m}=\mathcal{F}\left[a_{n}b_{n}e^{-i\omega_{m}2q_{n}/c}\right]_{m},
\label{transfer_func}
\end{equation}
where $H$ is the received signal transfer function and $\mathcal{F}[\cdot]$ is the discrete Fourier transform. Fig. \ref{qbin} shows how $q_{n}$ can be interpreted as the offset of the $n$th scatterer from the $n$th bin location where the bin widths are set to the canonical resolution limit. The number of bins is equal to the time bandwidth product of the transmit signal $N=TB$. This is the maximum number of scatterers that can be detected in a given ping.
\par
The evaluation of the Fourier transform in eq. \ref{transfer_func} is complicated by the exponential containing both a frequency component and an unknown offset $q_{n}$. In typical radar, sonar, and lidar systems, it is assumed that $q_{n}=0$. This would immediately simplify eq. \ref{transfer_func} and lead to the simple result that the measured impulse response function is just the scattering amplitudes $a_{n}b_{n}$ for each bin. Previous attempts to improve range resolution can be loosely interpreted as focusing on extracting extra ranging information from different parts of eq. \ref{transfer_func}. In \cite{Komissarov2019}, the signal phase was periodically shifted to improve the range estimate which corresponds to extracting range information from the exponential term. In \cite{Howell2023}, an ``amplitude reference" in the transmit waveform was used to estimate the target strength of the scatterers (expressed here as $a_{n}$). This was then used to determine the transmission loss and range which equates to estimating $q_{n}$ from the $b_{n}$ terms here.
\par
We believe that all approaches to solve the ranging problem are fundamentally equivalent to searching for either an exact or approximate solution to eq. \ref{transfer_func}. However, it is not yet clear how one would know which approach is best for a given application. A reasonable lower bound to ranging performance could be the canonical resolution limit. This is easily enforced by bounding the offsets such that $0\le q_{n}<\delta$; indicating that a solver for eq. \ref{transfer_func} should never perform worse than standard ranging methods. It is less obvious what an upper bound on performance would look like which would guide the selection of an ideal ranging algorithm for a given situation. However, all ranging algorithms described can only extract ranging information from the received signal. This means that if the total amount of ranging information in the return signal can be estimated, then an upper bound on range resolution can likewise be determined.

\section{Signal Information Content and Resolution}
Previous approaches to derive range resolution limits \cite{Richards2014, Komissarov2019, Jordan2023, Jordan2024} and the information content of electromagnetic signals \cite{hupfl2024} have relied on using the Fischer information to determine a lower bound on the variance of the range estimation. We choose instead to use a version of the Shannon information here. The calculation of Fischer information generally requires knowledge of both the particular transmit waveform used, and the processing steps applied to the received waveform, in order to determine the variance of the resulting range estimation. The Shannon approach typically ignores these details and assumes that there is some waveform and corresponding signal processing steps (normally described as a code) that will extract all the relevant information from the signal. This means we can find a general limit to the available signal information and hence range information which does not depend on a particular waveform or algorithm. This is especially useful for machine learning approaches \cite{Li2021} where those details may be unavailable.
\par
To begin, if a received signal $y$ is always a member of a finite set of $K$ possible signals $\left\{y_{k}\right\}$ each with a corresponding probability of occurring $\left\{p_{k}\right\}$, then the Shannon information $\mathcal{I}$ (in bits) gained per ping by receiving the $k$th signal is \cite{Brillouin2013},
\begin{equation}
\mathcal{I}_{k}=-\log_{2} p_{k}.
\label{inf_def}
\end{equation}
A finite set of possible signals $\left\{y_{k}\right\}$ can be constructed if we require that a noisy signal can always be uniquely mapped to one of the signals in the set. Note that the number of possible signals $K$ is different than the the number of basis functions $N$ since different scalings and combinations of the basis functions can result many unique signals $y_{k}$.
\par
\begin{figure}
\centering
\includegraphics[width=\columnwidth]{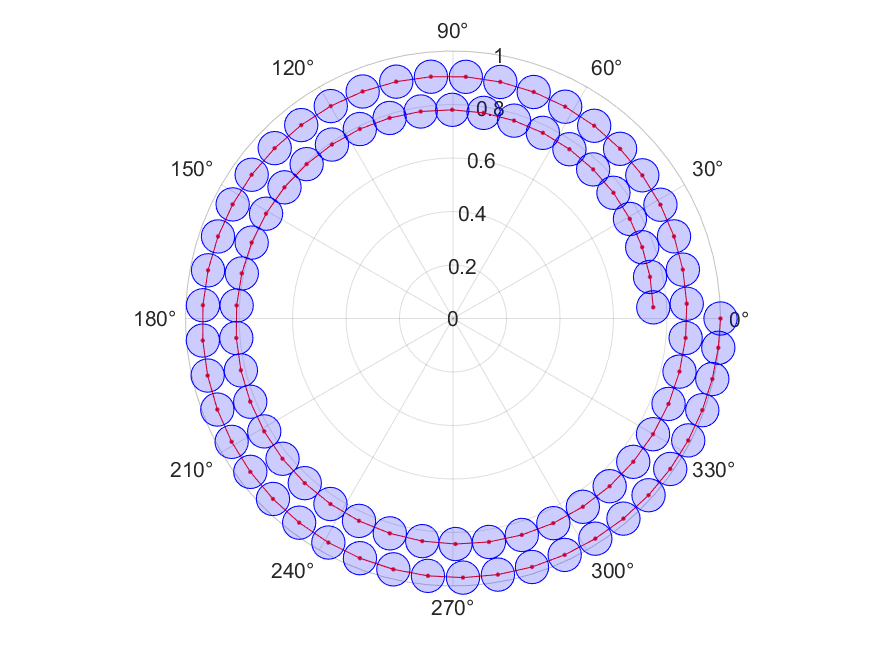}
\caption{The amplitude and phase of a single frequency as a function of scatterer range is shown by the solid spiral line. The number of uniquely identifiable ranges are illustrated by the dots at the center of the circles. A received signal falling within one of the circles is always mapped to the unique range corresponding to the center dot of that circle.}
\label{single_tone_spiral}
\end{figure}

To help visualize what these signal mappings would look like, we first consider the received signal from a single scatterer projected onto the I-Q plane for a given frequency. As the range of the scatterer increases, the amplitude of each frequency decreases and the phase is shifted due to the increase in propagation time. This is shown by the solid spiral line in fig. \ref{single_tone_spiral}. Noise obscures the exact position on the spiral of the scattered signal by nudging the received signal vector away from its original position on the spiral. Assuming additive white Gaussian noise, the ranging should be discretized so that a return that falls inside one of the blue circles, which has a radius equal to the square root of the noise vector magnitude, is mapped to its center red dot. In this way, the set of unique reference signals $\left\{y_{k}\right\}$ can be constructed.
\par
The maximum number of reference signals $K$ in the entire signal space is the number spheres with a radius $\sqrt{E_{n}}$, that can fit in a lager sphere with a radius $\sqrt{E_{s}+E_{n}}$, where  $E_{s}$ $(E_{n})$ is the total energy of the signal (noise). In this way, a signal vector in frequency space will remain uniquely identifiable, even when it is nudged away from its initial position by noise. This is similar to the approach for finding the channel capacity in communications systems \cite{Brillouin2013}. We are not specifically interested in how to partition this high dimensional space rather, we are looking for an absolute upper bound for the partitioning of the space. An upper bound to the number of smaller spheres that can fit inside a larger sphere is just the ratio of their volumes. In this case, that ratio is,
\begin{equation}
K\le\frac{\left(\sqrt{E_{s}+E_{n}}\right)^{2N}}{\sqrt{E_{n}}^{2N}}=\left(1+\gamma\right)^{N},
\end{equation}
where $\gamma=E_{s}/E_{n}$ is the SNR. Assuming that all possible signals $y_{k}$ are equally likely, $p_{k}=1/K$ and eq. \ref{inf_def} gives the upper bound to the total information content of the received ping,
\begin{equation}
\mathcal{I}=N\log_{2}\left(1+\gamma\right).
\end{equation}
\par
We are now in a position to determine how closely spaced in range the scatterers can be while still being resolvable. Let us consider the rescaling $q_{n}=n(\frac{c}{2B}-\delta)$ in eq. \ref{transfer_func}, which gives a uniform bin spacing of $\delta$. We discretize the distance $D$ covering the range of scatterers into $J$ bins such that the resolution is now $\delta=D/J$. The presence and target strength of a scatterer at a given range can be represented by one of $L$ discrete amplitude levels at each bin, where $L$ controls the fidelity of target strength measurements per bin. The total amount of information required to represent the scatterers in this way is,
\begin{equation}
\mathcal{I}=J\log_{2}\left(L\right).
\end{equation}
The information needed to represent the binned scatterers cannot exceed the information content of the signal. This leads to the inequality,
\begin{equation}
L^{J}\leq \left(1+\gamma\right)^{N}.
\label{bw_ineq}
\end{equation}
\par
This is equivalent to stating that the number of unique scatterer distributions that can be output from some ranging algorithm cannot be greater than the total number of possible unique received signals. In other words, we expect any ranging algorithm to be deterministic such that a given input can never map to more than one unique output. Eq. \ref{bw_ineq} can be put in terms of the bandwidth $B$ using the time bandwidth product.
\begin{equation}
L^{D/\delta}\leq \left(1+\gamma\right)^{TB}.
\end{equation}
The minimum resolvable bin width is now constrained by lower bound,
\begin{equation}
\delta\geq \frac{D}{TB}\frac{\log_{2}\left(L\right)}{\log_{2}\left(1+\gamma\right)},
\label{gen_res_limit}
\end{equation}
where $T$ is the length of the received signal or the time between pings. In many cases, there may be scatterers beyond the range $D$. To prevent the positions of faraway scatterers from aliasing to closer apparent positions, the maximum time between pings is $T=2D/c$. In this situation, Eq. \ref{gen_res_limit} can be written in terms of the canonical resolution bin width $\delta_{0}$ as,
\begin{equation}
\delta\geq \delta_{0}\frac{\log_{2}\left(L\right)}{\log_{2}\left(1+\gamma\right)}.
\label{res_limit}
\end{equation}
\par
Note that eq. \ref{res_limit} reduces to eq. \ref{canon_res_limit} when $L=1+\gamma$. In this case, all of the amplitude information for each ping goes into setting the amplitude level per bin. Therefore, the amplitude information per time bin $\log_{2}\left(L\right)$ equals the available information per frequency bin $\log_{2}\left(1+\gamma\right)$ and eq. \ref{res_limit} reduces to the canonical resolution limit. This is the way that the amplitude information is used in typical radar and sonar systems where the scatterer intensity is determined from the output of a matched filter.
\par
Eq. \ref{res_limit} suggests that it is possible to exceed the canonical resolution limit if the SNR is high enough. The highest resolution is obtained when $L=2$ such that each bin is a binary ``yes or no" of whether a scatterer is present in that bin. In this case, an SNR of $\gamma=3$ is sufficient to double the range resolution over the canonical limit. (Note that the SNR can also generally be improved by averaging the received signal over multiple pings.) The increase in resolution is possible since the signal amplitude and phase also contain ranging information. By utilizing this information for ranging rather than just assigning intensity and phase values for each bin, it is possible to achieve greater range resolution than is possible from using the output of a matched filter alone.
\begin{figure}
\begin{tikzpicture}
\fill[pattern=north east lines, pattern color=orange!40!gray] (1,1) rectangle ++(1.5,0.5);
\fill[pattern=north east lines, pattern color=orange!40!gray] (0.5,1.5) rectangle ++(2,0.5);
\fill[pattern=north east lines, pattern color=orange!40!gray] (1.5,2) rectangle ++(1.0,1.0);
\draw [line width=3] (1,1) -- ++(0, 0.5);
\draw (1,1) -- ++(1.5, 0);
\draw [line width=3] (0.5,1.5) -- ++(0, 0.5);
\draw (0.5,1.5) -- ++(0.5, 0);
\draw (0.5,2) -- ++(1, 0);
\draw [line width=3] (1.5,2) -- ++(0, 1);
\draw (1.5,3) -- ++(1, 0);
\draw (2.5,1) -- ++(0, 2);

\fill[cyan!50!gray!40] (-0.25,3.5) rectangle ++(1,0.33);
\draw[line width=1,draw=black] (-1.25,3.5) rectangle ++(1,1);
\draw[line width=1,draw=black] (-0.25,3.5) rectangle ++(1,1);
\draw[line width=1,draw=black,fill=cyan!50!gray!40] (0.75,3.5) rectangle ++(1,1);
\draw[line width=1,draw=black] (1.75,3.5) rectangle ++(1,1);

\draw[line width=1,draw=black] (-1.25,-0.5) rectangle ++(0.5,1);
\draw[line width=1,draw=black] (-0.75,-0.5) rectangle ++(0.5,1);
\draw[line width=1,draw=black] (-0.25,-0.5) rectangle ++(0.5,1);
\draw[line width=1,draw=black,fill=cyan!50!gray!40] (0.25,-0.5) rectangle ++(0.5,1);
\draw[line width=1,draw=black,fill=cyan!50!gray!40] (0.75,-0.5) rectangle ++(0.5,1);
\draw[line width=1,draw=black,fill=cyan!50!gray!40] (1.25,-0.5) rectangle ++(0.5,1);
\draw[line width=1,draw=black] (1.75,-0.5) rectangle ++(0.5,1);
\draw[line width=1,draw=black] (2.25,-0.5) rectangle ++(0.5,1);

\draw[dashed] (-1.25,3.83) -- ++(4,0);
\draw[dashed] (-1.25,4.17) -- ++(4,0);

\draw[] (-2,4) node[above] {$\delta=2$};
\draw[] (-2,4) node[below] {$L=4$};
\draw[] (-2,0) node[above] {$\delta=1$};
\draw[] (-2,0) node[below] {$L=2$};
\draw[] (-1.25,-0.5) node[below] {$0$};
\draw[] (-0.75,-0.5) node[below] {$1$};
\draw[] (-0.25,-0.5) node[below] {$2$};
\draw[] (0.25,-0.5) node[below] {$3$};
\draw[] (0.75,-0.5) node[below] {$4$};
\draw[] (1.25,-0.5) node[below] {$5$};
\draw[] (1.75,-0.5) node[below] {$6$};
\draw[] (2.25,-0.5) node[below] {$7$};
\draw (2.75,-0.5) -- ++(0.25, 0);
\draw (2.75,0.5) -- ++(0.25, 0);
\draw[] (3.25,-0.5) node[] {$0$};
\draw[] (3.25,0.5) node[] {$1$};

\draw[] (-1.25,4.5) node[above] {$0$};
\draw[] (-0.25,4.5) node[above] {$2$};
\draw[] (0.75,4.5) node[above] {$4$};
\draw[] (1.75,4.5) node[above] {$6$};
\draw (2.75,4.5) -- ++(0.25, 0);
\draw (2.75,3.83) -- ++(0.25, 0);
\draw (2.75,4.17) -- ++(0.25, 0);
\draw (2.75,3.5) -- ++(0.25, 0);
\draw[] (3.25,3.5) node[] {$0$};
\draw[] (3.4,3.83) node[] {$1/3$};
\draw[] (3.4,4.17) node[] {$2/3$};
\draw[] (3.25,4.5) node[] {$1$};

\draw[] (0.25,5.2) node[left] {$r$};
\draw[->] (0.25,5.2) -- (1.5,5.2);
\draw[->] (4.15,3.75) -- (4.15,4.25);
\draw[] (4.15,3.75) node[below] {TS};

\draw[] (-2.25,1.5) node[below] {Transceiver};
\draw[line width=1,draw=black,fill=gray!50] (-2.25,1.75) rectangle ++(0.5,0.5);
\node [trapezium, rotate=90, trapezium angle=50, minimum width=1.25cm, draw, thick, fill=gray!50] at (-1.5,2) {};
\draw[] (-1,1.5) arc (-30:30:1);
\draw[] (-0.75,1.25) arc (-30:30:1.5);
\end{tikzpicture}
\caption{The range is partitioned into discrete bins with width of either $\delta=1$ or $\delta=2$ (in arbitrary units) as shown. Each bin contains a normalized target strength (TS) value to indicate the presence of a scatterer. The smaller bin width is illustrated to only have two possible TS values, indicating a binary decision on the presence of a scatterer in a bin. The larger bin width allows for four TS measurements. The larger bin widths with more TS levels are better at capturing the differences in scattering area (bold surfaces) shown above, while the smaller bin widths are better at capturing the ranges of the different scattering surfaces.}
\label{target_binning}
\end{figure}
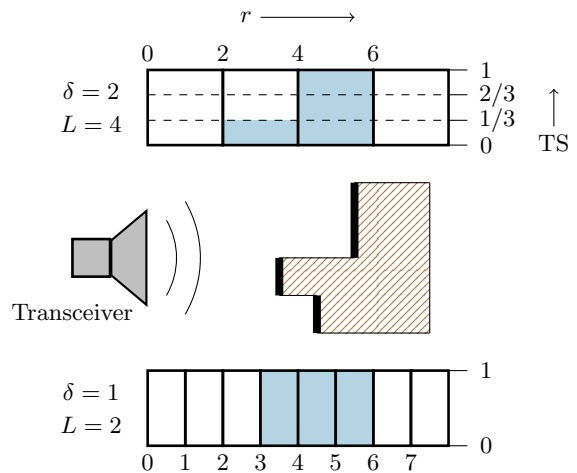
\par
More generally, eq. \ref{res_limit} gives a limit to how well a target can be detected and resolved for a given bandwidth and SNR; including for current systems that operate at the canonical resolution limit. Detection requires a non-zero amplitude for a given bin. To register a small but non-zero amplitude requires that $L$ is large, but this comes at the expense of range resolution and requires a larger bin size. On the other hand, high resolution of a target requires closely spaced bins. A smaller bin size however, requires a smaller value of $L$ which comes at the expense of target strength fidelity. Fig. \ref{target_binning} illustrates the trade-off on whether to direct the information from the signal to improve resolution or target strength measurements. Both measurements are often used for target identification in radar and sonar systems. Eq. \ref{res_limit} sets an upper bound on how much information is available to a given target identification algorithm and shows the connection between spatial resolution and target strength fidelity.

\section{Conclusion}
By applying ideas related to the Shannon view of information, a general lower bound to the bin size of any ranging algorithm is found. This lower bound is found to depend on the bandwidth, the SNR, and the target strength fidelity per bin. This limit can either be applied to systems operating at or above the canonical bin size to estimate the accuracy of target strength measurements or it can be applied to new approaches operating beyond the canonical limit to determine the smallest possible binning in range. Furthermore, our ideas also constrain the performance of any future machine learning approaches to the ranging problem since it is not necessary to know the processing details in deriving this new resolution limit.

\begin{acknowledgments}
This research was funded by the NSWC Panama City Division’s In-house Laboratory Independent  Research (ILIR) program from the Office of Naval Research (ONR).
\end{acknowledgments}

\appendix*

\section{Transmission Loss}
\label{transmission_loss}
For notational convenience, the time, distance, and frequency variables are scaled as $t=2B \bar{t}$, $r=2B\bar{r}/c$, \& $\omega=\bar{\omega}/2B$  where variables with an overbar are in their original units. The wave equation with viscous damping \cite{Tolstoy1923} can be put in the form,
\begin{equation}
\ddot{p}\left(\mathbf{r},t\right)-\nabla^{2}p\left(\mathbf{r},t\right)-\mu\nabla^{2}\dot{p}\left(\mathbf{r},t\right)=s\left(\mathbf{r},t\right),
\label{vis_wave_eq}
\end{equation}
where $p$ is the pressure, $s$ is the source term, and $\mu=\frac{8B}{3\rho_{w} c^{2}}\left(\mu_{s}+\frac{3}{4}\mu_{v}\right)$. Here, $\rho_{w}$ is the density of water and $\mu_{s}$ \&  $\mu_{v}$ are the shear and volume viscosities of water respectively. In the frequency domain, solutions to eq. \ref{vis_wave_eq} satisfy the inhomogeneous Helmholtz equation,
\begin{equation}
\left(\nabla^{2}+k^{2}\right)p\left(\mathbf{r},\omega\right)=-S_{k}\left(\mathbf{r}\right)
\label{helmholtz}
\end{equation}
where the wavenumber $k$ is related to $\omega$ and $\mu$ by,
\begin{equation}
\left(1-i\mu\omega\right)k^{2}+\omega^{2}=0.
\end{equation}
Expanding $k$ in terms of the small parameter $\mu$ produces the useful approximation,
\begin{equation}
k\approx\omega+i\frac{1}{2}\mu\omega^{2}.
\label{wave_num}
\end{equation}
\par
The Green's function of eq. \ref{helmholtz} for a point source located at $\mathbf{r}^{\prime}$ in free space takes the form \cite{Jackson1999},
\begin{equation}
G_{k}(\mathbf{r},\mathbf{r}^{\prime})=\frac{e^{ik\left|\mathbf{r}-\mathbf{r}^{\prime}\right|}}{4\pi\left|\mathbf{r}-\mathbf{r}^{\prime}\right|}.
\label{green_func}
\end{equation}
The general solution to eq. \ref{helmholtz} for a source distribution $S_{k}\left(\mathbf{r}^{\prime}\right)$ can be constructed using the Green's function $G_{k}\left(\mathbf{r},\mathbf{r}^{\prime}\right)$ according to the expression,
\begin{equation}
p_{k}\left(\mathbf{r}\right)=\int_{V}dV^{\prime}\;S_{k}\left(\mathbf{r}^{\prime}\right)G_{k}\left(\mathbf{r},\mathbf{r}^{\prime}\right).
\end{equation}
\par
The sound intensity $I$ is related to the peak pressure by $I=p^{2}/2\rho_{w} c$. The transmission loss in decibels can be calculated from eqs. \ref{wave_num} \& \ref{green_func} using the expression $\text{TL}=-10\log\left(I/I_{\circ}\right)$. Considering a reference intensity $I_{\circ}$ at a distance $r=1$ from the source, the transmission loss takes the form,
\begin{equation}
\text{TL}\left(r\right)=20\log(r)+\alpha\left(r-1\right),
\label{app_tl}
\end{equation}
where $\alpha=10\log(e)\mu\omega^{2}$ is the dimensionless absorption coefficient. It is well known however, that the absorption coefficient in seawater can be much higher than that predicted fom viscosity alone due to contributions from the dissociation and reassociation of \ce{MgSO4}. An expression for $\alpha$ including these other contributions is \cite{Urick1983},
\begin{equation}
\alpha=\frac{0.1f^{2}}{1+f^{2}}+\frac{40f^{2}}{4100+f^{2}}+2.75\times10^{-4}f^{2}+0.003,
\label{attenuation_eq}
\end{equation}
where $f$ is the frequency in kilohertz and $\alpha$ is given in units of $\text{dB}/\text{kiloyards}$.
\par
Note that eq. \ref{app_tl} is also valid for electromagnetic waves. The wave propagation for both wave types is given by eq. \ref{green_func} which controls the geometric spreading factor. The dissipation factor $\alpha$ is directly related to the imaginary part of the wavenumber $k$ which arises in electromagnetic waves due to dielectric loss.

\bibliography{ref}% Produces the bibliography via BibTeX.

\end{document}